\documentclass[iop]{emulateapj}
\usepackage{arydshln}
\slugcomment{{Accepted for publication in the Astrophysical Journal Letters}}

\begin{document}
\title{PSR~J1119--6127 and its pulsar wind nebula following the magnetar-like bursts}
\author{Harsha Blumer\altaffilmark{1,2}, Samar Safi-Harb\altaffilmark{3}, and Maura A. McLaughlin\altaffilmark{1,2}}
\altaffiltext{1}{Department of Physics and Astronomy, West Virginia University, Morgantown, WV 26506, USA; harsha.blumer@mail.wvu.edu}
\altaffiltext{2}{Center for Gravitational Waves and Cosmology, West Virginia University, Chestnut Ridge Research Building, Morgantown, WV 26505, USA; maura.mclaughlin@mail.wvu.edu}
\altaffiltext{3}{Department of Physics and Astronomy, University of Manitoba, Winnipeg, MB R3T 2N2, Canada; Samar.Safi-Harb@umanitoba.ca}

\begin{abstract}

We present a Chandra Director's Discretionary Time observation of PSR~J1119-6127 and its compact X-ray pulsar wind nebula (PWN) obtained on 27 October 2016, three months after the \textit{Fermi} and \textit{Swift} detection of millisecond bursts in hard X-rays, accompanied by $\gtrsim$160~times increase in flux. This magnetar-like activity, the first observed from a rotation-powered radio pulsar, provides an important probe of the physical processes that differentiate radio pulsars from magnetars. The post-burst X-ray spectrum of the pulsar can be described by a single powerlaw model with a photon index of 2.0$\pm$0.2 and an unabsorbed flux of 5.7$_{-1.1}^{+1.4}\times$10$^{-12}$~ergs~cm$^{-2}$~s$^{-1}$ in the 0.5--7.0~keV energy range. At the time of \textit{Chandra} observations, the pulsar was still brighter by a factor of $\sim$22 in comparison with its quiescence. The X-ray images reveal a nebula brighter than in the pre-burst Chandra observations (from 2002 and 2004), with an unabsorbed flux of 2.2$^{+1.1}_{-0.9}\times$10$^{-13}$ ergs~cm$^{-2}$~s$^{-1}$. This implies a current X-ray efficiency of $\approx$0.001 at a distance of 8.4~kpc. In addition, a faint torus-like structure is visible along the southeast-northwest direction and a jet-like feature perpendicular to the torus towards the southwest. The PWN is best fitted by an absorbed powerlaw with a photon index of 2.2$\pm$0.5 (post-burst). While the pulsar can still be energetically powered by rotation, the observed changes in PSR~J1119--6127 and its PWN following the magnetar-like bursts point to an additional source of energy powered by its high-magnetic field.
\end{abstract}


\section{Introduction}
\label{1}

The radio pulsar J1119--6127, at the center of the supernova remnant (SNR) G292.2--0.5, was discovered in the Parkes multibeam 1.4 GHz pulsar survey with a spin period $P$=408~ms, spin-down rate $\dot{P}$=4$\times$10$^{-12}$ s~s$^{-1}$, characteristic age $\tau_c$$\approx$1.9~kyr, and spin-down luminosity $\dot{E}$=2.3$\times$10$^{36}$~ergs~s$^{-1}$ (Camilo et al. 2000). Its surface dipole magnetic field strength is $B$=4.1$\times$10$^{13}$~G, close to the quantum critical field strength $B_{QED}$=4.4$\times$10$^{13}$~G, making PSR~J1119--6127 a high-magnetic field pulsar.  It has also shown sporadic, or rotating radio transient-like behavior, preceded by large amplitude glitch-induced changes in the spin-down parameters (Weltevrede et al. 2011).

The X-ray counterpart to the radio pulsar was first resolved with \textit{Chandra} in 2002, providing the evidence for a compact pulsar wind nebula (PWN) of $\sim$3$\arcsec$$\times$6$\arcsec$ in angular size (Gonzalez \& Safi-Harb 2003). A follow up study performed with \textit{Chandra} in 2004, combined with the 2002 observation, allowed for an imaging and spectroscopic study of the pulsar and PWN independently of each other (Safi-Harb \& Kumar 2008). The PWN showed elongated jet-like features extending at least $\sim$7$\arcsec$ north and south of the pulsar, and a longer southern jet becoming evident after accumulating $\sim$130~ks of combined \textit{Chandra} data. The pulsar spectrum was described by a two-component model consisting of a powerlaw (PL) with photon index $\Gamma$$\approx$1.9 and a thermal component, either a blackbody (BB) with temperature $kT$$\approx$0.21~keV or a neutron star atmospheric (NSA) model with $kT$$\approx$0.14~keV.  The PWN spectrum was described by a nonthermal model with $\Gamma$=1.1--1.4 (Safi-Harb \& Kumar 2008). An \textit{XMM-Newton} study of the pulsar revealed pulsations with an unusually high pulsed fraction of 74$\pm$14\% in the 0.5--2.0 keV energy range (Gonzalez et al. 2005). The profile is single-peaked and phase-aligned with its radio counterpart. No evidence of pulsations was detected above 2.5 keV (Ng et al. 2012).  Gamma-ray pulsations were reported from PSR~J1119--6127 using \textit{Fermi}, making it the source with the highest inferred $B$-field detected among $\gamma$-ray pulsars (Parent et al. 2011). 

On 2016 July 27 and 28, PSR J1119--6127 exhibited two short (0.02--0.04~s), energetic hard X-ray bursts detected by the \textit{Fermi} Gamma-ray Burst Monitor and \textit{Swift} Burst Alert Telescope (Younes et al. 2016; Kennea et al. 2016; G\"o\u{g}\"u\c{s} et al. 2016). Using the \textit{Swift} X-ray Telescope (XRT) and \textit{NuSTAR} data, Archibald et al. (2016) reported a large glitch ($\Delta$$\nu$/$\nu$=5.74(8)$\times$10$^{-6}$), pulsar flux increase by a factor $\gtrsim$160 (0.5--10~keV), and the appearance of strong X-ray pulsations above 2.5~keV for the first time. On the other hand, the pulsed radio emission from PSR~J1119--6127 disappeared after the onset of magnetar-like bursts, but reappeared two weeks later (Burgay et al. 2016a, 2016b; Majid et al. 2017). Using the \textit{Fermi} Large Area Telescope data obtained from 2016 July 27--August 12, Younes et al. (2016) reported the lack of detection of $\gamma$-ray pulsations and derived an upper limit, consistent with the measured pre-burst $\gamma$-ray flux of the pulsar.

\begin{figure*}[ht]
\includegraphics[width=0.5\textwidth]{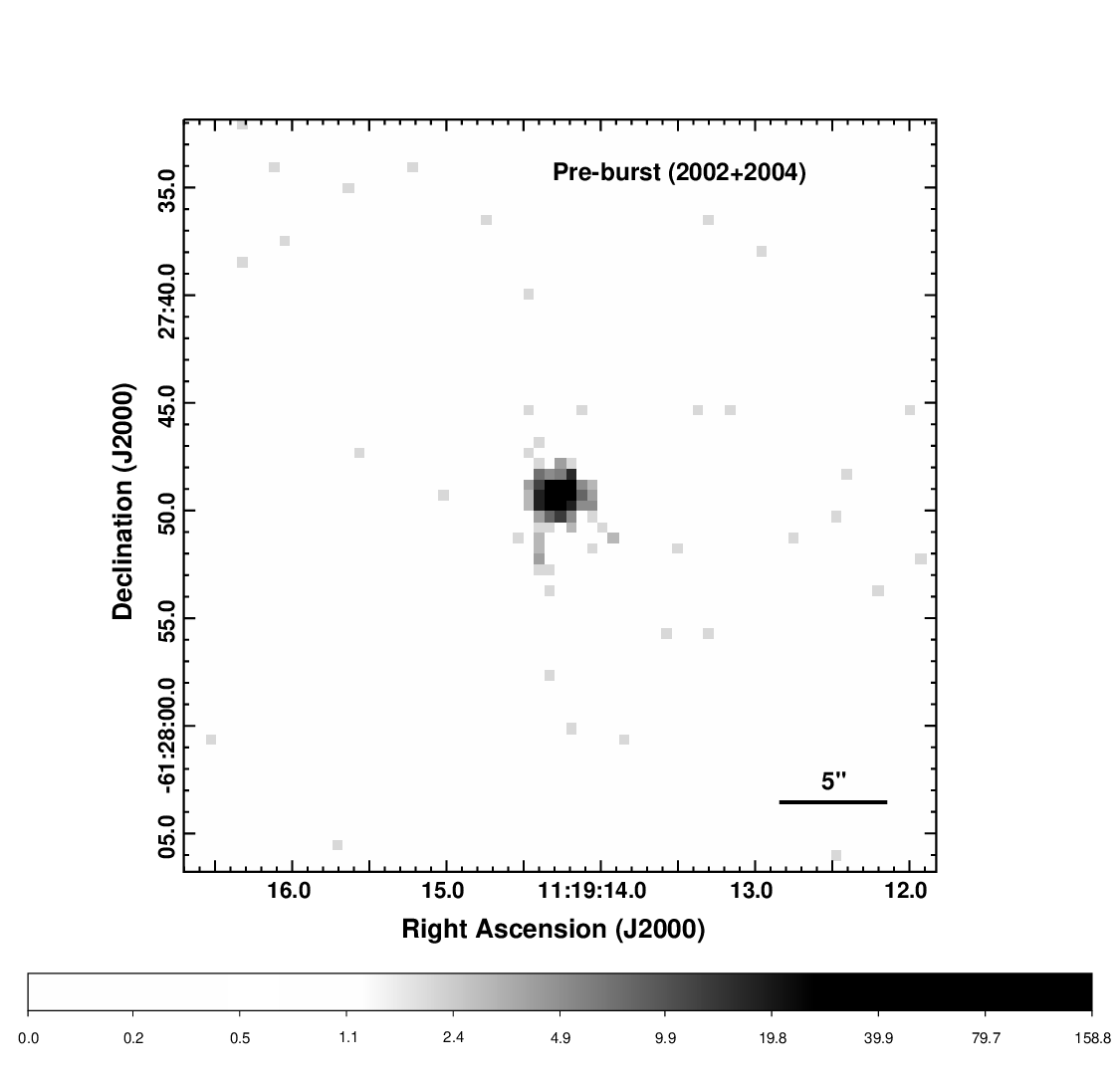}
\includegraphics[width=0.5\textwidth]{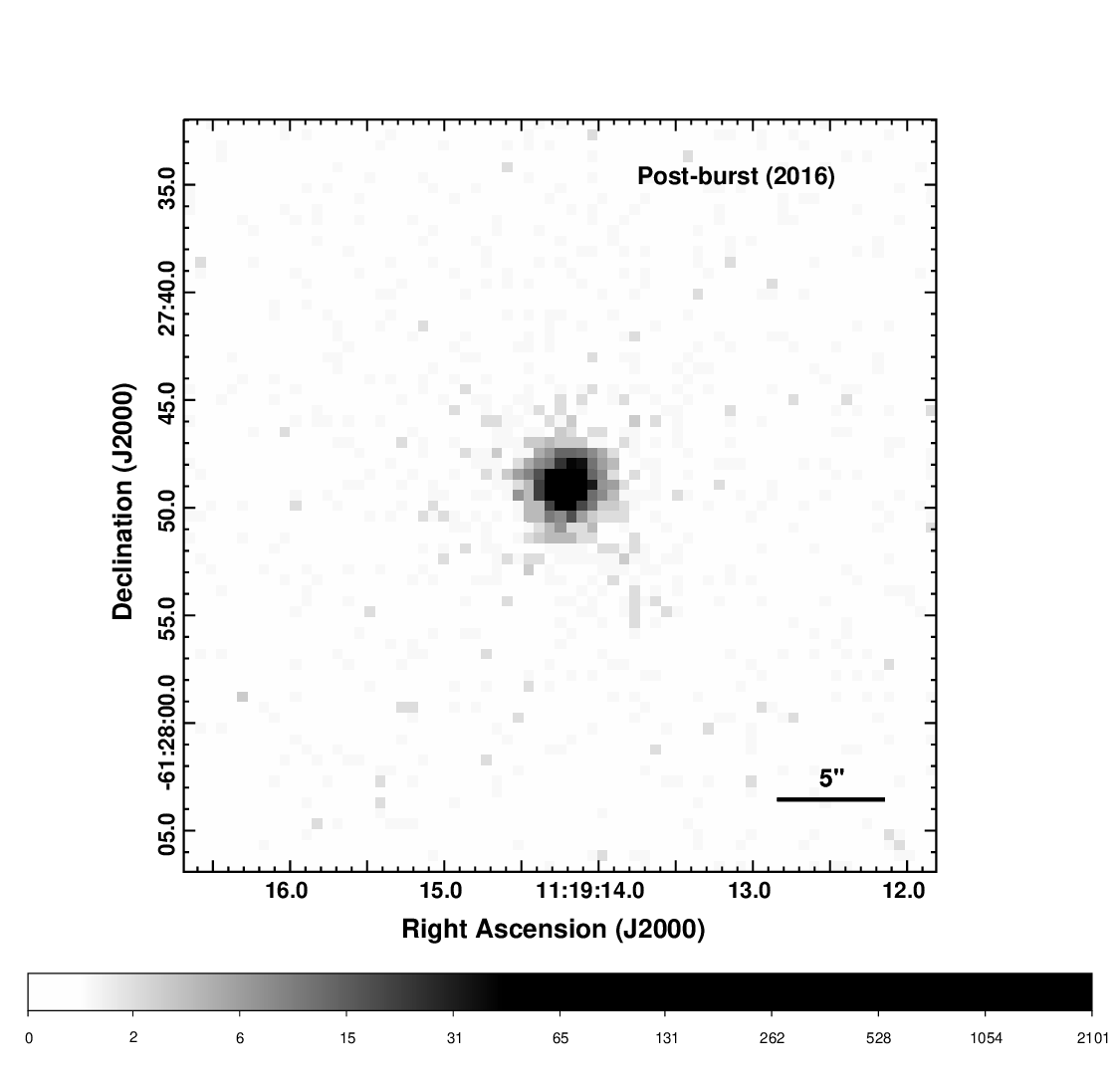}
\includegraphics[width=0.5\textwidth]{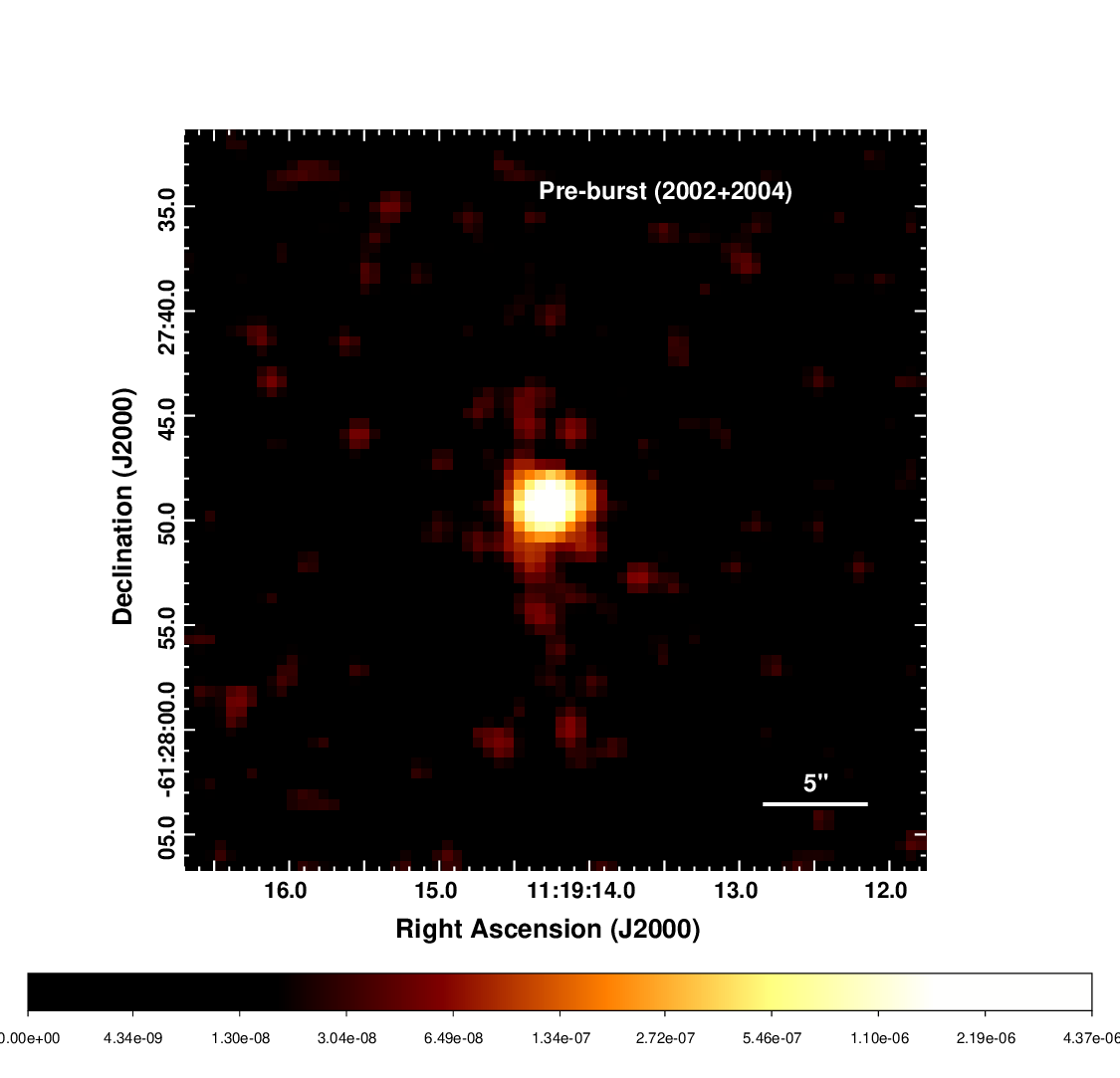}
\includegraphics[width=0.5\textwidth]{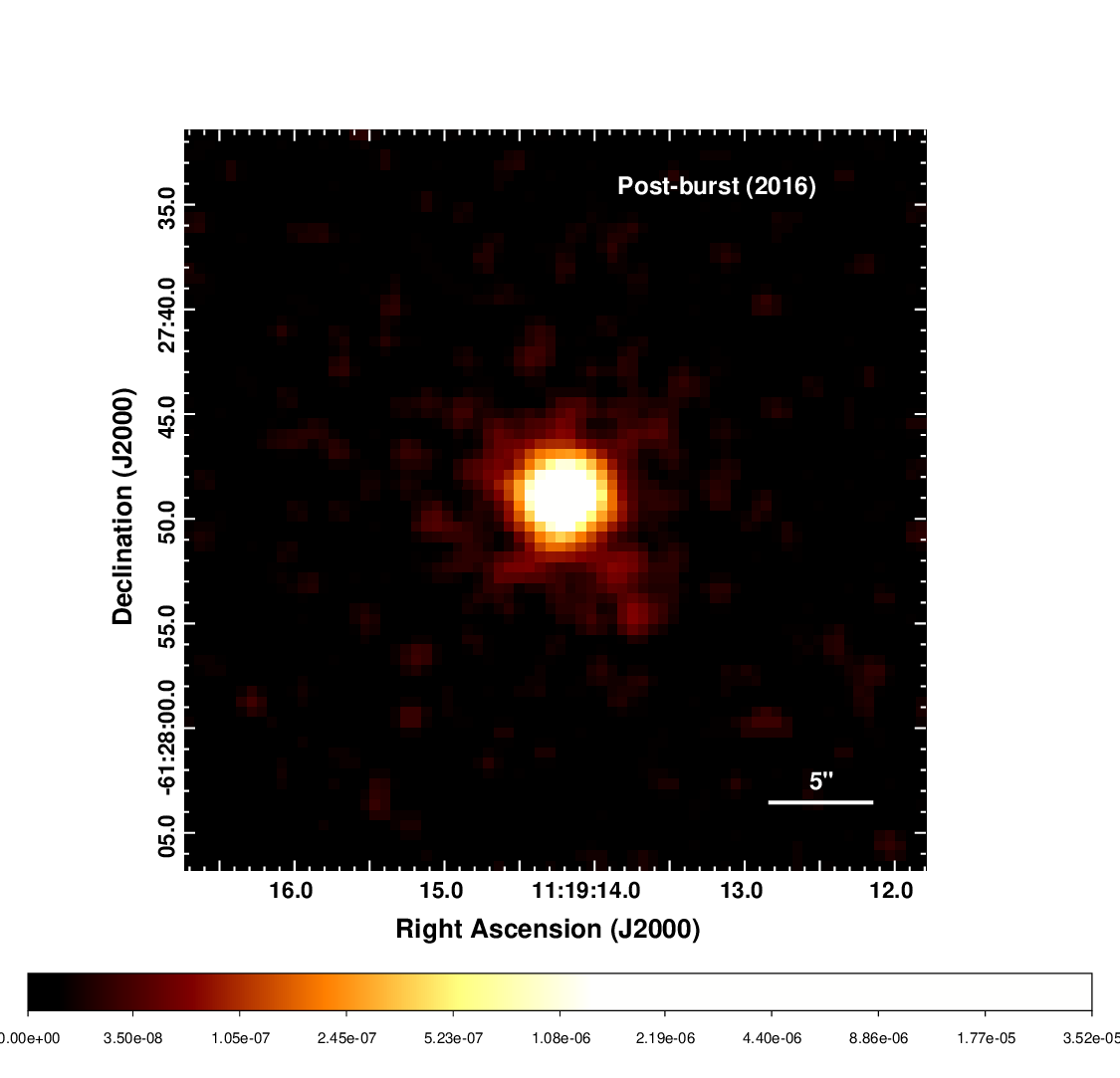}
\caption{Broadband (0.5--7.0~keV) pre-burst (left) and post-burst (right) images of PSR~J1119--6127 and its PWN in logarithmic scale. The top and bottom panels show the raw, unsmoothed and exposure corrected, smoothed images in units of counts~pixel$^{-1}$ and photons~cm$^{-2}$~s$^{-1}$~arcsec$^{-2}$, respectively. The pre-burst image was obtained by combining the 2002 and 2004 observations.  The post-burst image shows faint nebulosity near the saturated PSF wings of the bright pulsar that suggests jet and tori-like features, consistent with a PWN interpretation. North is up and East is to the left. }
\end{figure*}

We requested a \textit{Chandra} Director's Discretionary Time observation to study the pulsar and, particularly, the effect of the magnetar-like bursts on its surrounding nebula. The superior angular resolution of \textit{Chandra} is required to disentangle the emission of the pulsar and its compact PWN. Here, we present these results together with a reanalysis of the archived pre-burst \textit{Chandra} observations. The distance to the PSR~J1119--6127 is taken as 8.4 kpc from HI absorption measurements to the SNR (Caswell et al. 2004) and we scale all derived quantities in units of $d_{8.4}$=$D/8.4$~kpc, where $D$ is the actual distance to the pulsar.

\begin{figure*}[th]
\includegraphics[width=0.5\textwidth]{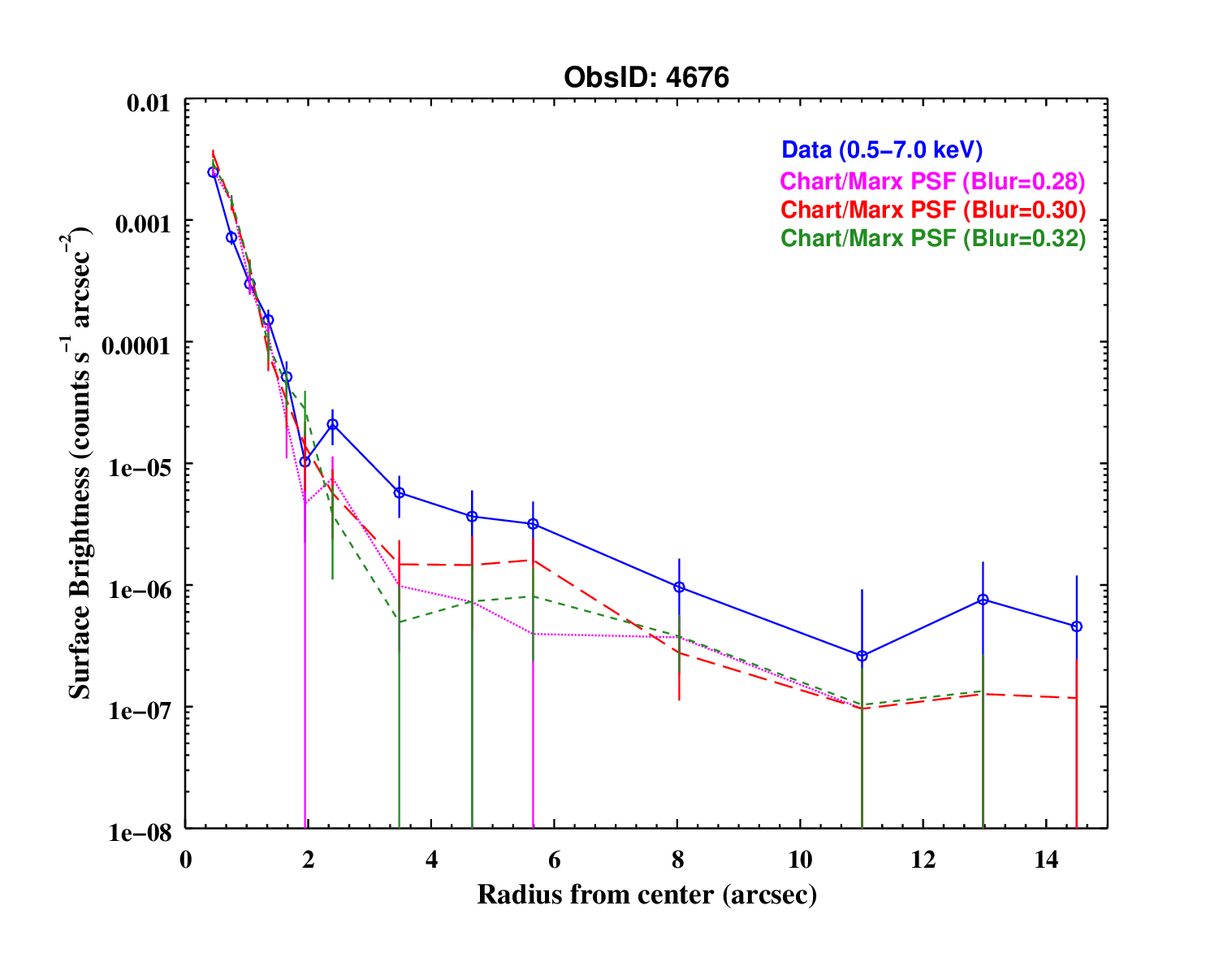}
\includegraphics[width=0.5\textwidth]{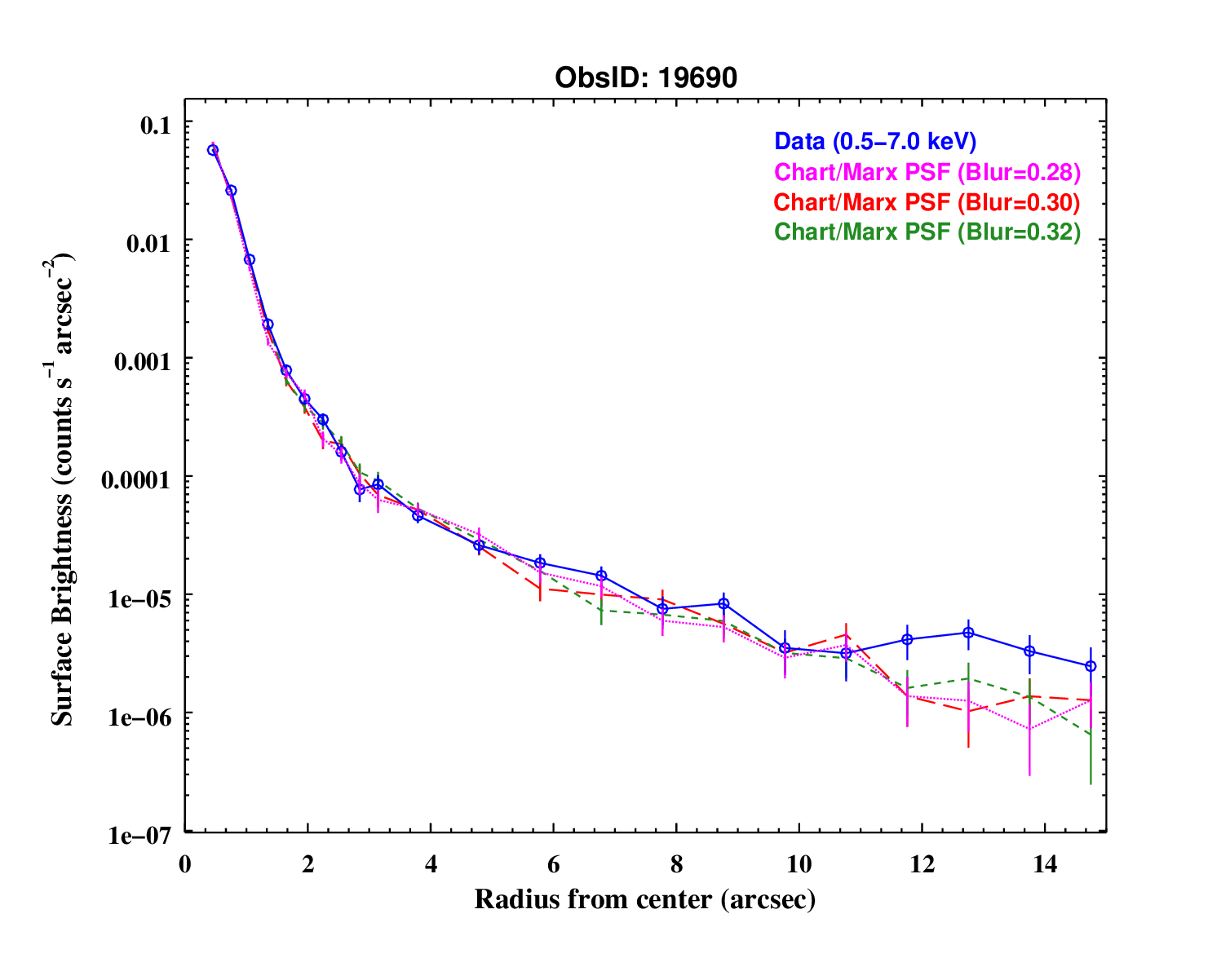}
\caption{Surface brightness profiles of PSR~J1119--6127 for different blur values. The broadband (0.5--7.0 keV) radial profiles are shown for one of the pre-burst (left) and post-burst (right) observations. }
\end{figure*}

\section{Observation and Data Reduction}
\label{2}

We obtained new observations of PSR~J1119$-$6127 on 2016 October 27 (ObsID 19690), three months after the reports of the magnetar-like bursts. We reprocessed all previous observations taken on 2004 November 2--3 (ObsID 6153), 2004 October 31--November 1 (ObsID 4676), and 2002 March 31 (ObsID 2833).  The target was positioned at the aimpoint of the Advanced CCD Imaging Spectrometer (ACIS) in all four observations. The standard processing of the data was performed using the \textit{chandra\_repro} script in CIAO version 4.9\footnote{http://cxc.harvard.edu/ciao} (CALDB v.4.7.6) to create new level 2 event files.  The resulting effective exposure times were 55.5~ks, 18.9~ks,  60.5~ks, and 56.8~ks for ObsIDs 19690, 6153, 4676, and 2833, respectively. Throughout this paper, we refer to the 2002 and 2004 \textit{Chandra} observations as `pre-burst' and the data taken in 2016 as `post-burst'.

\section{Imaging analysis}
\label{3.1}

Figure~1 shows the broadband (0.5--7.0 keV) pre-burst (left) and post-burst (right) zoomed-in raw, unsmoothed (top) and exposure corrected, smoothed (bottom) images of the region around PSR~J1119--6127, centered on the pulsar coordinates at $\alpha_{J2000}$=11$^{h}$19$^{m}$14\fs260 and $\delta_{J2000}$=$-$61\degr27\arcmin49\farcs30.  The bottom-panel images have been exposure-corrected using the CIAO task \textit{fluximage} with a binsize of 1 pixel and smoothed using a Gaussian function of radius 2 pixels. The pre-burst image shows a nebula of size $\sim$6$\arcsec$$\times$15$\arcsec$ in the north-south direction (Safi-Harb \& Kumar 2008) while the post-burst image shows small-scale fine structures around the pulsar. The post-burst PWN morphology along the southeast--northwest direction can be considered as an equatorial torus $\approx$10\farcs0$\times$2\farcs5  in size, running perpendicular to the small jet-like structure $\approx$1\farcs5$\times$3\farcs5 southwest of the pulsar. The overall PWN is titled at an angle of $\sim$35$\degr$--40$\degr$ towards the west. A detailed spatially resolved imaging or spectroscopic study of the PWN structures is not possible with the current data due to the paucity of photon counts.

To further constrain the morphology of the diffuse emission around the pulsar at small angular scales and to probe the pulsar extent, we used the \textit{Chandra} Ray Tracer (ChaRT\footnote{http://cxc.harvard.edu/ciao/PSFs/chart2/index.html}) and MARX\footnote{http://space.mit.edu/CXC/MARX/} (ver. 5.3.2) software packages. The point spread functions (PSFs) were simulated using the best-fit pulsar spectrum (see Section 4) and ChaRT for each observation. The ChaRT output was then supplied to MARX to produce the simulated event files and PSF images. Different values (0.25, 0.28, 0.30, 0.32, 0.35) of the \textit{AspectBlur} parameter (which accounts for the known uncertainty in the determination of the aspect solution) were used to search for an excess corresponding to the extended emission. We created broadband (0.5--7.0~keV) radial profiles up to 15$\arcsec$ by extracting net counts in circular annuli centered on the point source, with an annular background from 30$\arcsec$--40$\arcsec$, and rebinned the data to obtain better statistical precision. Figure~2 shows a comparison of the surface brightness profiles for the pre-burst (ObsID 4676) and post-burst (ObsID 19690) data with the profiles generated with ChaRT/MARX for different blur values. The data show some slight excess compared to the simulated profiles beyond a radius of $\approx$1\farcs5 for both the pre-burst and post-burst observations, although excess is clearly seen only beyond 11$\arcsec$ for the post-burst data.

We performed spatial fitting on the 0.5--7.0~keV pulsar image to study the morphology of the extended emission quantitatively with \textit{Sherpa}\footnote{http://cxc.harvard.edu/sherpa/}. The ChaRT/MARX generated PSF was loaded as a table model to be used as a convolution kernel for the point source emission. The multi-component source model in \textit{Sherpa} included a 2D Beta model (\textit{beta2d}) for describing the extended component of the source emission, a PSF-convolved 2D Gaussian model (\textit{gauss2d}) to describe the point source component, and a \textit{const2d} model to describe the constant background level contributing to the total emission. The best-fit parameters were determined by the C-statistic and Nelder-Mead optimization method. We obtained a diffuse emission radius of 10\farcs.3$\pm$1\farcs.2 and 6\farcs.2$\pm$0\farcs.8 for the post-burst and pre-burst data,  respectively, in comparison with the point source full-width half maximum (FWHM) of 1\farcs2, suggesting an expansion of the nebula.

\begin{figure*}[th]
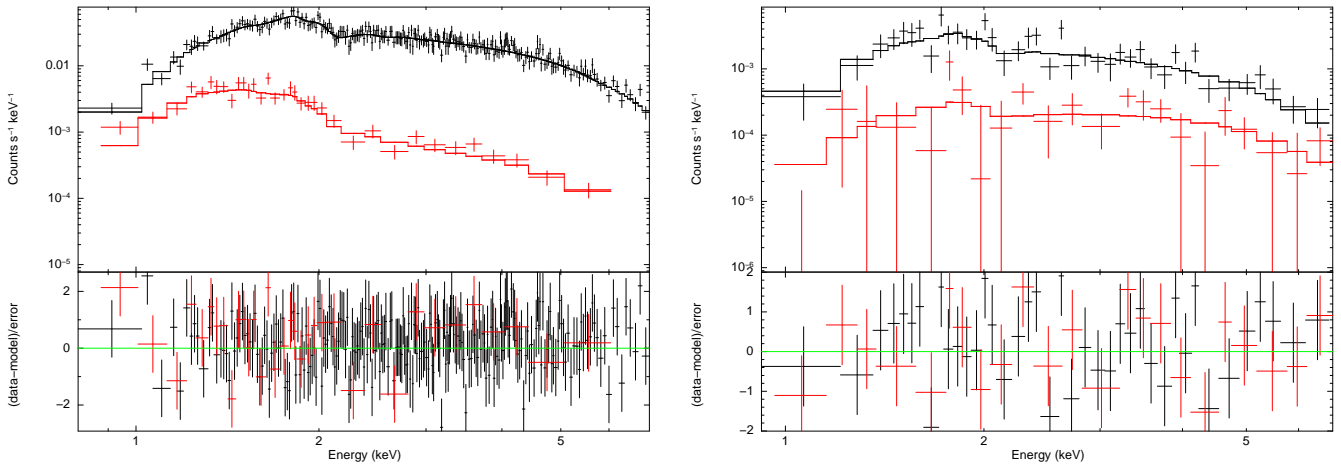

\includegraphics[angle=-90, width=0.5\textwidth]{f3a.eps}
\includegraphics[angle=-90, width=0.5\textwidth]{f3b.eps}
\caption{{Left}: Post-burst (black; PL fit) and pre-burst (red; BB+PL fit) X-ray spectra of PSR~J1119--6127.  {Right}: Post-burst (black) and pre-burst (red) PWN spectra fitted with a PL model.  The bottom panel shows the residuals in terms of sigmas. The pre-burst data are combined together using the \textit{combine\_spectra} task in CIAO and rebinned for display purposes.}
\end{figure*}

\section{Spectral analysis}
\label{4}

The spectral analysis was performed using the X-ray spectral fitting package, XSPEC version 12.9.1, and restricted to 0.5--7.0~keV as these energy bands were not background dominated. The contributions from background point sources were removed prior to the extraction of spectra.  All the spectra extracted were grouped by a minimum of 10 counts per bin and the errors were calculated at the 90\% confidence level. 

\subsection{Pulsar spectrum}

For a point source observed on-axis with \textit{Chandra}, $\sim$90\% of the encircled energy lies within 1\farcs2 at 1.49~keV and within 2\farcs5 at 6.4~keV\footnote{http://cxc.harvard.edu/proposer/POG/html/chap6.html}.  Guided by the radial profile plots (Figure~2), 90\% encircled energy fraction, and sherpa modelling, we here consider 1\farcs5 as the best extraction radius for the pulsar. We chose an annular ring of 3$\arcsec$--5$\arcsec$ centered on the pulsar as background, to minimize the contamination from the surrounding PWN. Due to the evidence of pulsar brightening in the post-burst data, we investigated the possibility of pileup using the CIAO task \textit{pileup\_map}\footnote{http://cxc.harvard.edu/ciao/ahelp/pileup\_map.html.}. We obtained an average of 0.2~photons per frame of 3.2~s in the centermost pixel of the post-burst pulsar image, which translates into a pileup fraction of $\sim$10\%. For a quantitative estimate of the post-burst pulsar spectrum, we used the \textit{jdpileup} model of the \textit{Chandra} spectral fitting software \textit{Sherpa} convolved with an absorbed PL and BB model, which gave a pileup fraction of 5.2\% and 21\% for the two models, respectively. The pre-burst data did not show any evidence of pileup. 

The post-burst pulsar spectrum was first fit with different one- and two-component models.  A PL model yielded a better fit ($\chi^2_{\nu}$/dof=0.906/291) with $N_H$=1.7$\times$10$^{22}$~cm$^{-2}$ and $\Gamma$=1.8, while the BB model gave a low $N_H$=0.7$\times$10$^{22}$~cm$^{-2}$ and $kT$=1.0~keV for $\chi^2_{\nu}$/dof=1.437/291, with excess emission seen above $\sim$3~keV.  The addition of a second component was statistically not required for a single PL model, but the fit improved when a PL component was added to the BB model with $N_H$=1.6$^{+0.3}_{-0.2}$$\times$10$^{22}$~cm$^{-2}$, $kT$=0.4$\pm$0.1~keV, and $\Gamma$=1.5$^{+0.3}_{-0.4}$ for $\chi^2_{\nu}$/dof=0.892/289. We next fitted the spectra by including a pileup component to the PL and BB models, and the results are shown in Table~1. The pre-burst spectrum of the pulsar was also explored with different models, and as elaborated in Safi-Harb \& Kumar (2008), a two-component BB+PL model best described the pulsar spectra (Table~1). Figure~3 (left) shows the best-fit post-burst (PL; black) and pre-burst (BB+PL; red) pulsar spectra. To better evaluate the contamination in the spectra from the surrounding PWN, the pulsar spectra were explored using larger extraction radii of 2\farcs0 and 2\farcs5 and we obtained similar spectral parameters as for the 1\farcs5 region. 

\subsection{PWN spectrum}

We extracted an annular ring of 2$\arcsec$--10$\arcsec$ region for the overall PWN to determine any spectral variations in the PWN between the epochs, following the imaging analysis and sherpa modelling results. The background was extracted from a nearby source-free elliptical region\footnote{The spectral fits were explored with different annular and elliptical backgrounds and binning, and the spectral parameters were consistent within uncertainties to those shown in Table 1.}.  All the extracted regions were simultaneously fit with a PL model by tying $N_H$ together. Figure~3 (right) and Table~1 show the best-fit model and spectral parameters for the PWN. The pre-burst spectral fit results obtained here are consistent with the results for the $\sim$6$\arcsec$$\times$15$\arcsec$ region presented in Safi-Harb \& Kumar (2008). 

\begin{table*}[ht]
\caption{Spectral fits to the PSR J1119--6127 and its PWN}
\center
\begin{tabular}{l l l l l l l}
\hline\hline
Parameter & \multicolumn{3}{c}{Pulsar} &  \multicolumn{2}{c}{PWN} \\
\cline{2-6}
 & Pre-burst & \multicolumn{2}{c}{Post-burst} & Pre-burst & Post-burst \\
 \cline{3-4}
& BB+PL & pileup*PL & pileup*BB & PL & PL \\
\hline
$ N_{H}$ ($10^{22}$ cm$^{-2}$) & 1.6$^{+0.7}_{-0.5}$ & 1.8$\pm$0.2 & 1.1$\pm$0.1 & \multicolumn{2}{c}{1.8$^{+0.6}_{-0.5}$} \\
\cdashline{5-6}
$\Gamma$ & 2.0$_{-0.9}^{+0.8}$ & 2.0$\pm$0.2 & \nodata  &  1.2$\pm$0.8 & 2.2$\pm$0.5 \\

kT (keV) &  0.2$\pm$0.1 & \nodata & 0.7$\pm$0.1 &  \nodata & \nodata\\

$F_{unabs}$ (PL)$^a$ & 7.4$_{-1.8}^{+1.0}\times$10$^{-14}$ & 5.7$_{-1.1}^{+1.4}\times$10$^{-12}$ & \nodata & 2.3$_{-1.5}^{+3.5}\times$10$^{-14}$ & 2.2$_{-0.9}^{+1.1}\times$10$^{-13}$ \\

$F_{unabs}$ (BB)$^a$ & 1.8$_{-0.8}^{+1.5}\times$10$^{-13}$ &  \nodata & 1.1$^{+0.3}_{-0.1}\times$10$^{-11}$  &  \nodata & \nodata \\

$\chi_{\nu}^2$/dof & 1.008/60 & 0.893/289 & 0.988/289 & \multicolumn{2}{c}{0.912/57}  \\
\cdashline{5-6}
Count rate$^b$ & (4.8$\pm$0.3)$\times$10$^{-3}$ & \multicolumn{2}{c}{(1.08$\pm$0.01)$\times$10$^{-1}$} & (8.3$\pm$2.3)$\times$10$^{-4}$ & (6.9$\pm$0.4)$\times$10$^{-3}$ \\
\cdashline{3-4}

$L_{X}$$^c$ & 2.2$_{-1.1}^{+1.4}\times$10$^{33}$ & 4.8$_{-0.9}^{+1.2}\times$10$^{34}$ & 9.3$_{-0.8}^{+2.5}$$\times$10$^{34}$ & 1.9$_{-1.3}^{+1.9}\times$10$^{32}$ & 1.9$_{-0.8}^{+0.9}\times$10$^{33}$  \\

$\eta_X$=$L_{X}$/$\dot{E}$ & 0.001 & 0.02 & 0.04 & 0.0001 & 0.001 \\

\hline
\end{tabular}
\tablecomments{Galactic absorption is modeled with \textit{tbabs} in XSPEC (Wilms et al. 2000). The post-burst pileup fractions for the PL and BB models are 5.2\% and 21\%, respectively. The pre-burst and post-burst PWN spectra are simultaneously fit by tying their $N_H$ together.  Errors are at 90\% confidence level. \\
$^a$ Unabsorbed flux (0.5--7.0 keV) in units of ergs~cm$^{-2}$~s$^{-1}$. \\
$^b$ Background subtracted count rates (0.5--7.0 keV) in units of counts~s$^{-1}$. \\
$^c$ X-ray luminosity (0.5--7.0 keV) in units of ergs s$^{-1}$ assuming isotropic emission at a distance of 8.4~kpc. \\
 }
\end{table*}

\section{Discussion and conclusion}
\label{4}

The high-magnetic field ($B$$\gtrsim$$B_{QED}$) pulsars are believed to be an important class of objects for studying the relationship between magnetars and radio pulsars. Seven high-$B$ pulsars have been identified so far. These include the radio pulsars J1119--6127, J1718--3718, J1734--3333, J1814--1744, J1847--0130, the X-ray pulsar J1846--0258, and the rotating radio transient J1819--1458 (Ng \& Kaspi 2011). Based on the X-ray observations of PSR~J1718--3718, Kaspi \& McLaughlin (2005) suggested that high-$B$ pulsars may be quiescent magnetars. The first evidence for such a link was found when PSR~J1846$-$0258 in SNR Kes~75 showed magnetar-like bursts and radiative changes such as a flux increase and spectral softening in X-rays (Gavriil et al. 2008; Kumar \& Safi-Harb 2008). PSR~J1119--6127 is the first radio pulsar, and the second high-$B$ pulsar, to display a magnetar-like burst. Here, we present a discussion of the results from our pre-burst and post-burst study on PSR J1119--6127 and its compact nebula.

The \textit{Chandra} observations of PSR~J1119--6127, made three months after its magnetar-like bursts, can be described by a single PL ($\Gamma$=2.0$\pm$0.2) or BB ($kT$=0.7$\pm$0.1~keV) model with pileup, in contrast to its quiescent spectrum, which required a combination of PL and BB models. Here, we prefer a PL model over a BB model for the post-burst pulsar data due to the following reasons. Firstly, the pileup fraction required to fit the spectrum with a BB model is much higher than a PL model, suggesting that a BB fit is only possible if a large pileup fraction absorbs high-energy photons that are better fit with a PL tail. Secondly, the $N_H$ obtained from a BB model is much lower compared to that obtained for the pulsar and its PWN (Table~1), as well as from the SNR diffuse emission regions immediately surrounding the pulsar (Kumar et al. 2012). Thirdly, when the post-burst pulsar spectrum was fitted with a BB+PL model (although a second component was not statistically required), we find that $\sim$85\% of the total unabsorbed flux is dominated by the nonthermal component. These results, together with the fact that the pulsar's emission beyond 3~keV could not be described by a BB model alone, imply that the X-ray emission in the post-burst state is mainly magnetospheric in nature.  Assuming isotropic emission, the pulsar's luminosity $L_{X,PSR}$=4$\pi$$d_{8.4}^2$$F_{unabs}$$\approx$4.8$\times$10$^{34}$~$d_{8.4}^2$~ergs~s$^{-1}$, implying an X-ray efficiency $\eta_{X, PSR}$=$L_{X, PSR}/\dot{E}$$\approx$0.02 in the 0.5--7.0~keV energy range. It is interesting to note that the pulsar's $\eta_X$ is less than 1 during its magnetar-like burst, indicating that its spin-down energy could still power the X-ray emission.  Magnetar bursts are usually accompanied by dramatic changes in the persistent emission properties and spectral evolution, such as hardening/softening, change in pulsed fraction, pulse profiles, flux changes etc. (Rea \& Esposito 2011). The burst-induced radiative changes observed for PSR~J11119--6127 are very similar to those seen in magnetars, suggesting an activity associated with the pulsar's high-magnetic field.  Such results have been found in the case of PSR~J1846--0258 as well, further implying that the high-$B$ pulsars could be powered by both rotational and magnetic energy (Camilo 2008).  

The pulsar's pre-burst data showed a compact PWN of size $\sim$6$\arcsec$$\times$15$\arcsec$, while we see a change in the PWN morphology with faint tori and jet-like features surrounding the pulsar in the post-burst data.  In the 92 days that elapsed from the detection of the first burst to the Chandra observation, the maximum distance the ejected particles could have traveled, assuming an 8.4 kpc distance and a speed of light velocity, is 1\farcs9. Therefore, the new extended feature of $\sim$10$\arcsec$ radius cannot be associated with the recent burst unless the distance to the pulsar is overestimated by at least an order of magnitude. The PWN spectrum also showed a change in photon index from 1.2$\pm$0.8 to 2.2$\pm$0.5 following the burst, although not unusual since the X-ray spectra of most PWNe have $\Gamma$=1--2.5 (Kargaltsev \& Pavlov 2008). The post-burst X-ray luminosity of the PWN is 1.9$_{-0.8}^{+0.9}\times$10$^{33}$~ergs~s$^{-1}$ (0.5--7.0~keV), implying an X-ray efficiency $\eta_{X,PWN}$$\approx$0.001, consistent with the typical values of $\sim$10$^{-5}$ to 10$^{-1}$ observed in other rotation-powered pulsars with PWNe (Kargaltsev \& Pavlov 2008). The flux from the compact nebula has also increased by an order of magnitude in comparison with the pulsar's quiescent state, consistent with the pre-burst and post-burst surface brightness profiles (Figures~1 and 2). Although small-scale variabilities in PWN structures are seen in many nebulae, the changes as observed in PSR~J1119--6127 are difficult to interpret in terms of spin-down energy alone. 

\begin{table*}[th]
\caption{Comparison of the X-ray properties of PWNe around high-$B$ pulsars and magnetars}
\center
\begin{tabular}{l l l l l l l l l l}
\hline\hline
Pulsar & P & $B$  & Distance & $\dot{E}$ & $\Gamma$ & $L_{X}$/$\dot{E}$ & Ref. \\
 & (s) & (10$^{13}$~G) & (kpc) & (ergs~s$^{-1}$) & & & & \\
 \hline
PSR~J1119--6127$^a$ & 0.408 & 4.1 & 8.4 & 2.3$\times$10$^{36}$  & 2.0$\pm$0.2 & 0.001 & This work \\
PSR~J1819--1458 & 4.26 & 5.0 & 3.6 & 3.0$\times$10$^{32}$  & 3.7$\pm$0.3  & 0.15 & Rea et al. 2009; Camero-Arranz et al. 2013\\
Swift~J1834.9--0846 & 2.48 & 14 & 4.0 & 2.1$\times$10$^{34}$  & 2.2$\pm$0.2 & 0.1 & Younes et al. 2016 \\
PSR~J1846--0258 & 0.324 & 4.9 & 6.0 & 8.1$\times$10$^{36}$  & 1.93$\pm$0.03 & 0.02  & Kumar \& Safi-Harb 2008; Ng et al. 2008 \\
SGR~J1935+2154$^b$ & 3.24 & 22 & 11.7 & 1.7$\times$10$^{34}$ & 3.8$\pm$0.3 & 0.35  & Israel et al. 2016; Surnis et al. 2016\\
\hline
\end{tabular}
\tablecomments{$^a$ Post-burst X-ray efficiency is quoted here (pre-burst $\eta_X$$\sim$0.0001; Safi-Harb \& Kumar 2008).\\
$^b$ Diffuse emission could be either a dust scattering halo or a wind nebula. \\
 }
\end{table*}

It has been proposed that magnetars can produce relativistic particle outflows during an outburst or from a steady flux of Alfv\'en waves powering a wind nebula (Thompson \& Duncan 1996; Harding 1999). Signatures of X-ray emission from a wind nebula have been reported in a few high-$B$ pulsars and magnetars such as PSR~J1846--0258, PSR~J1819--1458, Swift~J1834.9--0846, and SGR~J1935+2154 (Safi-Harb 2013). Table~2 summarizes the X-ray properties of PWNe observed around these highly magnetized neutron stars.  PSR~J1846--0258, which features a very prominent X-ray nebula, showed small-scale variability in its PWN after its magnetar-like bursts in 2006 (Ng et al. 2008; Kumar \& Safi-Harb 2008). For the extended emission around the PSR~J1819--1458 (Rea et al. 2009; Camero-Arranz et al. 2013), Swift~J1834.9--0846 (Kargaltsev et al. 2012; Esposito et al. 2013), and SGR~J1935+2154 (Israel et al. 2016), the authors favor a PWN or a scattering halo origin. Bright X-ray sources with large column densities can lead to an extended dust scattering halo, with the halo brightness proportional to the source flux (Predehl \& Schmitt 1995). The diffuse emission region around Swift~J1834.9--0846 has been identified as an inner symmetric region ($\lesssim$50$\arcsec$) of scattering halo and an outer asymmetric region ($\sim$150$\arcsec$) of a possible magnetar wind nebula, since the emission remained fairly constant in flux and spectral shape across three years (Younes et al. 2012, 2016). 

A detailed investigation of a halo component associated with PSR~J1119--6127's magnetar-like burst would require modeling beyond the scope of this Letter. However, for a dust scattering halo, one would expect to find symmetric structures around the point source with a relatively steeper (softer) photon index than the source as the scattering cross-section varies with the inverse-square of the energy. The fact that the compact PWN has an asymmetric structure (Figure~1), with a hard spectral index (2.2$\pm$0.5) comparable to that of the pulsar (2.0$\pm$0.2) does support a PWN interpretation. We further note that the spectral index of PSR~J1119--6127 is also harder than the other magnetars associated with dust scattering haloes. Israel et al. (2016) suggests that the extended emission seen around SGR~1935+2154, with a $\Gamma$=3.8$\pm$0.3, could also be magnetically powered due to its unusually high X-ray efficiency. From Table~2, we see that the X-ray efficiency is much less than 1 for PSR~J1119--6127, but its photon index is comparable to the magnetically powered nebula around Swift~J1834.9--0846 ($\Gamma$=2.2$\pm$0.2). Despite some similarities in properties, none of other PWNe exhibited any notable change in morphology or spectrum as seen for PSR~J1119--6127.

In summary, while the PWN in PSR~J1119--6127 can still be energetically powered by its spin-down power, the changes observed in the PWN spectrum point towards a new source of energy powered by the magnetar bursts. The change in PWN morphology could be somewhat related to the recent bursts, but the large scale changes must have happened over longer timescales and could perhaps be related to an earlier undetected burst. Unfortunately, the spacing between observations is insufficient to make firm conclusions about the timescale of these changes.  We cannot rule out a dust scattering halo component for PSR J1119--6127, but require additional deeper observations at different epochs to separate any halo component and confirm the nature of the PWN emission and its morphology post-burst.

\acknowledgments
The authors are grateful to Patrick Slane and the \textit{Chandra} science team for making this DDT observation possible, as well as thank the \textit{Chandra} HelpDesk for assistance with data analysis. We thank the referee for a very careful reading that helped improve and clarify the manuscript. HB and MAM are supported by the NSF award number 1516512. SSH acknowledges support by the Natural Sciences and Engineering Research Council of Canada (through the Canada Research Chair and Discovery Grant programs) and the Canadian Space Agency. This research made use of NASA's ADS and HEASARC maintained at the Goddard Space Flight Center.

\end{document}